\documentstyle[psfig,prc,aps]{revtex}
\begin{document}
\draft
\title{Additional $J/\Psi$ Suppression from High Density Effects}
 \author{M.B. Gay Ducati
$^{1,\dag}$\footnotetext{$^{\dag}$ E-mail:gay@if.ufrgs.br}, V.P.
Gon\c{c}alves $^{2,*}$\footnotetext{$^{*}$
E-mail:barros@ufpel.edu.br}, L. F. Mackedanz
$^{1,\star}$\footnotetext{$^{\star}$
E-mail:thunder@if.ufrgs.br} }
\address{$^1$ Instituto de F\'{\i}sica, Universidade Federal do Rio Grande do
Sul\\
Caixa Postal 15051, CEP 91501-970, Porto Alegre, RS, BRAZIL}
\address{$^2$ Instituto de F\'{\i}sica e Matem\'atica,  Universidade
Federal de Pelotas\\
Caixa Postal 354, CEP 96010-090, Pelotas, RS, BRAZIL}
 \maketitle

\begin{abstract}
 At  high energies the saturation effects associated to the high parton density should modify the
behavior of the observables in proton-nucleus and nucleus-nucleus
scattering. In this paper we investigate the saturation effects in
the nuclear $J/\Psi$ production and estimate the modifications in
the energy dependence of the cross section as well as in the
length of the nuclear medium.  In particular, we calculate the
ratio of $J/\Psi$ to Drell-Yan cross sections and show that it is
strongly modified if the high density effects are included.
Moreover, our results are compared with the data from
the NA50 Collaboration and predictions for the RHIC and LHC
kinematic regions are presented. We predict an additional $J/\Psi$
suppression associated to the high density effects.

\end{abstract}


\section{Introduction}
High energy heavy-ion collisions offer the opportunity  to study
the properties of the predicted QCD phase transition to a locally
deconfined quark-gluon plasma (QGP) \cite{rep}. A dense parton
system is expected to be formed in the early stage of relativistic
heavy-ion collisions at RHIC (Relativistic Heavy Ion Collider)
energies and above, due to the onset of hard and semihard parton
scatterings. The search for experimental evidence of this
transition during the very early stage of  $AA$ reactions requires
to extract unambiguous characteristic signals that survive the
complex evolution through the later stages of the collision. One
of the proposed signatures of the QCD phase transition is the
suppression of quarkonium production, particularly of the $J/\psi
$ \cite{satz}. The idea of suppression of $c\overline{c}$ mesons
$J/\psi $, $\psi^{,}$, etc., is based on the notion that
$c\overline{c}$ are produced mainly via primary hard collisions of
energetic gluons during the preequilibrium stage up to shortly
after the plasma formation (before the initial temperature drops
below the production threshold), and the mesons formed from these
pairs may subsequently experience deconfinement when traversing
the region of the plasma. In a QGP, the suppression occurs due to
the shielding of the $c\overline{c}$ binding potential by color
screening, leading to the breakup of the resonance. The
$c\overline{c}$ ($J/\psi, \psi ^{\prime },...$) and
$b\overline{b}$ ($\Upsilon,\Upsilon^{\prime},...$) resonances have
smaller radii than light-quark hadrons and therefore higher
temperatures are needed to dissociate these quarkonium states.

Over the years, several groups have measured the $J/\Psi$  yield
in heavy ion collisions with the $J/\Psi$ suppression observed in
the experimental data. In particular, the NA50 Collaboration at
CERN observed a much stronger $J/\Psi$ suppression in Pb-Pb
collisions at SPS energies \cite{abreu}. Different mechanisms have
been proposed to explain this phenomenon. It has been suggested
that the suppression was due  the QGP phase \cite{blaizot},
percolation deconfinement \cite{nardisatz}  or absorption by
comovers \cite{kharsatz,capella} (For a review see e.g. Ref.
\cite{review}). Shortly, the origin of the anomalous behavior of
the $J/\Psi$ production cross section is still debated and several
competing interpretations have so far been proposed. In general,
these models consider the final state interactions of the
quarkonium state with the nuclear/QGP medium. A comprehensive analysis
of the different mechanisms is presented in Ref. \cite{zpc74}. Here, we address
other search of $J/\Psi$ suppression in nuclear collisions at high
energies: the modification in the nuclear wave functions of the
incident nuclei associated to the high density (saturation)
effects.

The probability of a thermalized QGP production and the  resulting
strength of its signatures strongly depends on the initial
conditions associated to the distributions of partons in the
nuclear wave functions. At very high energies, the growth of
parton distributions should saturate. There is a possible
formation of a Color Glass Condensate \cite{iancu},  characterized
by a bulk momentum scale $Q_s$. Moreover, the limitation on the
maximum phase-space parton density that can be reached in the
hadron wave function (parton saturation) and very high values of
the QCD field strength squared $F^2_{\mu \nu} \propto 1/\alpha_s$
\cite{muesat} are expected in this regime. Furthermore, the number
of gluons per unit phase space volume practically saturates and at
large densities grows only very slowly (logarithmically) as a
function of the energy \cite{vicsat}. If the saturation scale is
larger than the QCD scale $\Lambda _{QCD}$, then this system can
be studied using weak coupling methods. The magnitude of $Q_s$ is
associated to the behavior of the gluon distribution at high
energies, and some estimates have been obtained. In general, the
predictions are $Q_s\sim 1$ GeV at RHIC and $Q_s\sim 2-3$ GeV at
LHC \cite{gyusat,vicslope}. From the experimental point of view,
the recent results from HERA for $ep$ collisions
\cite{golec,eskolaqiu} and from RHIC for heavy ion collisions
\cite{levin,mclraju,rajukra} suggest that these processes  at high
energies probe QCD in the non-linear regime of high parton
density. These results motivate our analysis of the $J/\Psi$
production in nuclear processes.

%One important point for the studies of $J/\Psi$ production  is
%that in the $J/\Psi$ cross section calculation using the collinear
%factorization framework we integrate over the momentum fraction of
%the incoming partons. It implies that we must know the behavior of
%the nuclear parton distributions in the full $x$ range to obtain
%realistic predictions for  these observable. The current estimates
%for the nuclear $J/\Psi$ production  (See e.g. Ref.
%\cite{ramonaklein}) consider as input the EKS parameterization for
%the nuclear parton distributions \cite{eks}, which are solutions
%of the DGLAP evolution equations \cite{dglap}. Consequently, these
%analysis do not consider the possible presence of  high density
%effects in the RHIC and LHC kinematical regions.  Therefore, in
%order to investigate the presence and magnitude of the high
%density effects in the nuclear $J/\Psi$ production considering the
%collinear factorization of the cross section, it is necessary to
%include these effects, as well as the antishadowing, EMC and Fermi
%motion effects, in the nuclear parton distributions. Here we use
%as input in our calculations the AG parameterization proposed in
%Ref. \cite{ayavic}, which improves the EKS one  by the inclusion
%of the perturbative high density effects \cite{ayala1}. This
%parameterization  deals with these effects using the
%Glauber-Mueller formula  which is the simplest one that reflects
%the main qualitative features of a more general approach based on
%non-linear evolution \cite{iancu,kovchegov}.
In this work, we analyze in detail the $J/\Psi$ production in $pA$ and $AA$
processes. A detailed study of the $J/\Psi$ production in $pA$
collisions is justified by the fact that a systematic study of
$pA$ and $AA$  scatterings  at the same energies  is essential to
gain insight into the structure of the dense medium effects. Such
effects, as the energy loss and high density effects, are absent
or small in $pp$ collisions, but become increasingly prominent in
$pA$ collisions, and are of major importance in $AA$ reactions. By
comparing $pA$ and $AA$ reactions involving very heavy nuclei, one
may be able to distinguish basic hadronic effects that dominate
the dynamics in $pA$ collisions, from a quark-gluon formation
predicted to occur in heavy ion $AA$ collisions. Furthermore, our
analysis is motivated by the fact that the quarkonium  production
at RHIC and LHC energies is dominated by initial state gluons.  As
the probability for making a heavy quark pair is proportional to
the square of gluon distribution, any depletion in number of
gluons will make a significant difference in the number of the
$J/\Psi$ produced \cite{epgaygc}.
Here we assume the presence of  initial state effects  in the nuclear wave
functions and  final state interactions of the $c \overline{c}$
pair with the nuclear medium. Our main goal is to estimate the magnitude
of quarkonium suppression in hadronic matter associated to these effects. Our
results demonstrate that the high density effects implies an
additional $J/\Psi$ suppression when compared with the scenario where
only  final state interactions are estimated. As the formation of a QGP is not assumed, our predictions are a
lower bound for the  $J/\Psi$ suppression in high energy nuclear collisions.

This paper is organized as follows. In the next  section, we
present the color evaporation model (CEM), which is used as a
model for the $J/\Psi$ production in proton-nucleus ($pA$) and
nucleus-nucleus ($AA$) collisions. In the section \ref{ISeffect} we
present a brief review of the AG parameterization \cite{ayavic}, used to include the
high density effects in the calculation. Moreover, we present our predictions for
the energy dependence of the $J/\Psi$ cross section in section \ref{firstresult}. In section
\ref{FSeffect} the generalization of the CEM to include the final
state interactions proposed in Ref. \cite{qvz} is discussed and
our results for the medium length dependence of the cross sections
are presented in section \ref{comments}. Our main conclusions are also
summarized.

\section{$J/\Psi$ production in the collinear factorization}
\label{jpsiprod}

Vector-meson production
has proven to be a very interesting process in which to test the
interplay between the perturbative and nonperturbative regimes of
QCD (For a review, see for example \cite{caldwell,schu}).
One of the main uncertainties in  quarkonium production
is related to the transition  from the colored state to a
colorless meson. Initially, the $q\overline{q}$ pair will in
general be in a color octet state. It subsequently neutralizes its
color and binds  into a physical resonance. Color neutralization
occurs by interaction with the surrounding color field.
An alternative view of the $J/\Psi$ production process is to use the color evaporation model
(CEM), which describes a large range of data in hadro- and
photoproduction, as shown in Refs. \cite{cemmodel}.
  In CEM, quarkonium production is treated identically to open
heavy quark production with exception that in the case of
quarkonium, the invariant mass of the heavy quark pair is
restricted to be below the open meson threshold, which is twice
the mass of the lowest meson mass  that can be formed with the
heavy quark. For charmonium  the upper limit on the
$c\overline{c}$ mass is then $2m_D$. The hadronization of the charmonium states from the
$c\overline{c}$ pairs is nonperturbative, usually involving the
emission of one or more soft gluons. Depending on the quantum
numbers of the initial $c\overline{c}$ pair and the final state
charmonium, a different matrix element is needed for the
production of the charmonium state. The average of these
nonperturbative matrix elements are combined into the universal
factor $F[nJ^{PC}]$, which is process- and kinematics-independent
and describes the probability that the $c\overline{c}$ pair binds
to form a quarkonium $J/\Psi (nJ^{PC})$ of given spin $J$,
parity $P$, and a charge conjugation $C$. Once $F$ has been fixed
for each state ($J/\Psi $  or $\Psi ^{\prime }$) the model
successfully predicts the energy and
momentum dependence \cite{vogt,satz2}.

Considering the  $J/\Psi$ production and the collinear
factorization approach, the CEM predicts that the cross section
in the collision of hadrons $A$ and $B$ is given by
\begin{eqnarray}
&&\sigma_{AB\to {\rm J}/\psi X}=
K \, \sum_{a,b}
\int d q^2 \left ( \frac{
\hat \sigma_{ab\to c \bar c}(Q^2)}{Q^2}\right )
\int dx_F \phi_{a/A}(x_a)\phi_{b/B}(x_b)
\frac{x_a x_b}{x_a+x_b}F_{c\bar c\to {\rm J}/\psi}(q^2),
\label{cross}
\end{eqnarray}
where $\sum_{a,b}$ runs over all parton flavors,
 $Q^2=q^2+4 m_c^2$, $\phi_{a/A}(x_a)$ is the distribution function
of parton $a$ in hadron $A$,
$x_F=x_a-x_b$ and $x_a x_b=Q^2/s \equiv \tau$. The expressions of the
elementary partonic cross sections can be taken from
 Ref. \cite{combridge}, and to take into account the next-to-leading
order corrections to the cross section we consider the phenomenological
constant $K$-factor. The factor $F_{c\bar c\to {\rm J}/\psi}(q^2)$
describes the transition probability for the $c\bar c$ state of the
relative square momentum $q^2$ to evolve into a physical $J/\psi$ meson.
In general, it is parameterized by
\begin{eqnarray}
F_{c\bar c\to {\rm J}/\psi}(q^2)
&=&N_{{\rm J}/\psi}\theta(q^2)\theta(4{m_D}^2-4m_c^2-q^2) \,\, ,
\label{fcem}
\end{eqnarray}
where  $N_{J/\Psi}$ is a normalization factor which is obtained
from a fit of the experimental data for $J/\Psi$ production in
proton-proton collisions. Here, we assume that $K\, N_{J/\Psi} =
0.250$, similarly to what is made in Refs. \cite{qvz,cemmodel},
where this model was successfully compared with the experimental
results.
For nuclear collisions, a number of experiments has measured a less than
linear $A$ dependence for various processes of production \cite{experiment,alfass}, which
indicates that the nuclear medium effects cannot be disregarded. 

\section{High Density Effects in $J/\Psi$ Production}
\label{ISeffect}

One important point for the studies of $J/\Psi$ production  is
that in the $J/\Psi$ cross section calculation using the collinear
factorization framework we integrate over the momentum fraction of
the incoming partons. It implies that we must know the behavior of
the nuclear parton distributions in the full $x$ range to obtain
realistic predictions for  these observable. The current estimates
for the nuclear $J/\Psi$ production  (See e.g. Ref.
\cite{ramonaklein}) consider as input the EKS parameterization for
the nuclear parton distributions \cite{eks}, which are solutions
of the DGLAP evolution equations \cite{dglap}. Consequently, these
analysis do not consider the possible presence of  high density
effects which should modify the dynamical evolution of the nuclear
parton distributions in the  RHIC and LHC kinematical regions.

As our focus is to analyze the influence of the high density effects in
the $J/\Psi$ production, here we only consider the medium
modifications in the nuclear parton distributions (nuclear
shadowing effect), disregarding the contributions of energy loss
and the intrinsic heavy-quark components for the non-linear $A$
dependence of the cross sections (For a discussion of these
effects in charmonium production see Ref. \cite{vogtxf}). The
nuclear shadowing is the modification of  the nuclear  parton
distributions so that $\phi_{a/A}(x, Q^2)\, \neq \phi_{a/p}(x,Q^2)$, 
as expected from a superposition of $pp$ interactions. The current 
experimental data presenting nuclear shadowing can be described 
reasonably using the DGLAP evolution equations \cite{dglap} with 
adjusted initial parton distributions \cite{eks}. However, this 
parameterization does not include the dynamical saturation effects 
predicted to modify the evolution equation and, consequently, the
behavior of the nuclear parton distributions  at small $x$ and large $A$ \cite{muesat}. In this kinematical 
regime, the density of quarks and gluons becomes very high and the 
processes of interaction and recombination between partons, not 
present in the DGLAP evolution, should be considered. 
%In Ref. \cite{ayavic} a procedure to improve 
%the nuclear parton distributions and include the perturbative high 
%density effects in these distributions was proposed. 
Therefore, in order to investigate the presence and magnitude of the high
density effects in the nuclear $J/\Psi$ production considering the
collinear factorization of the cross section, it is necessary to
include these effects, as well as the antishadowing, EMC and Fermi
motion effects, in the nuclear parton distributions. Here we use
as input in our calculations the AG parameterization proposed in
Ref. \cite{ayavic}, which improves the EKS one  by the inclusion
of the perturbative high density effects \cite{ayala1}. This
parameterization  deals with these effects using the
Glauber-Mueller formula  which is the simplest one that reflects
the main qualitative features of a more general approach based on
non-linear evolution \cite{iancu,kovchegov}.
For completeness, we now present a qualitative discussion of the main properties of the
approach used by this parameterization.

\begin{figure}[t]
\centerline{\psfig{file=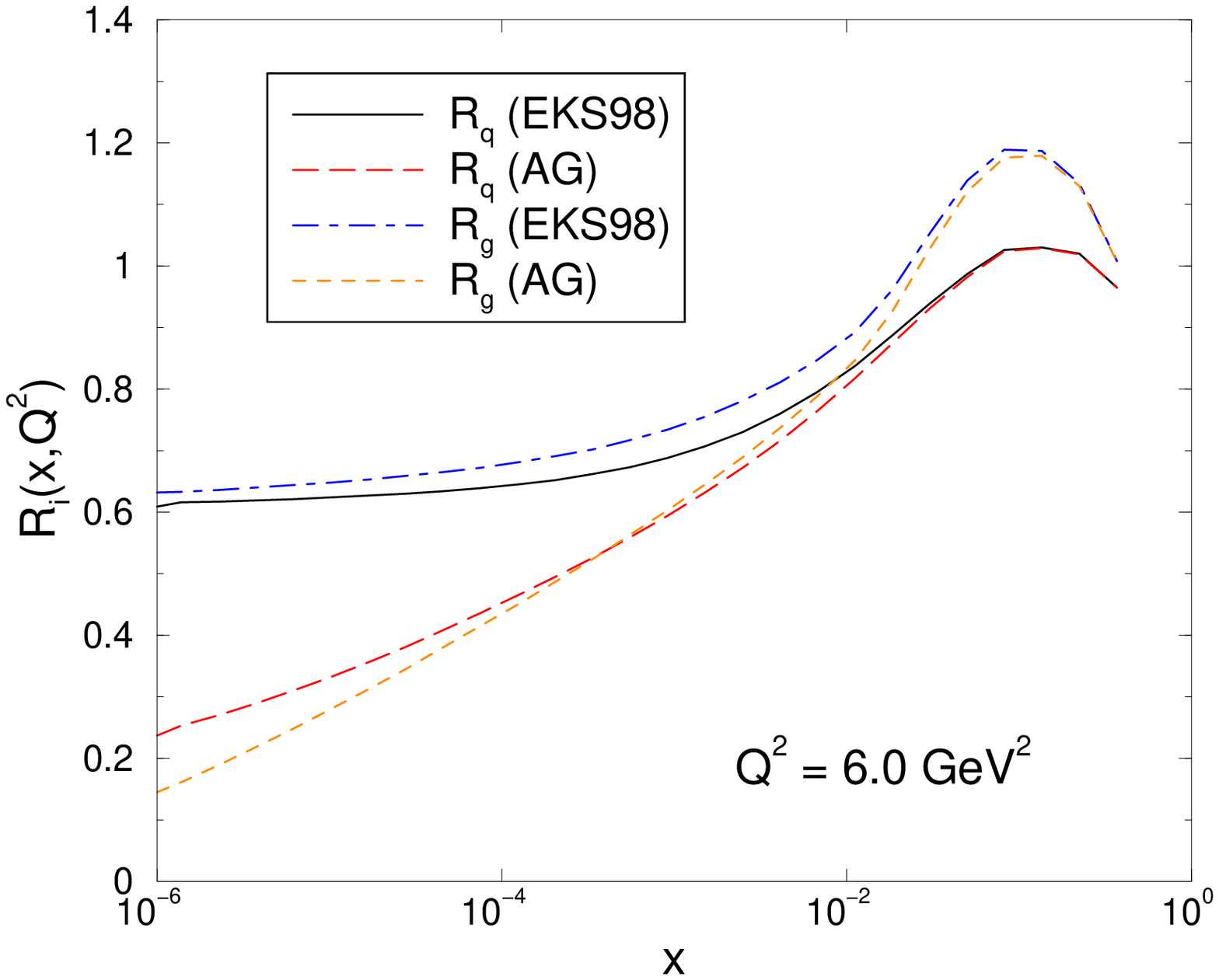,width=150mm}} \caption{The $x$-dependence of the nuclear
ratios $R_{g}$ and $R_{q}$ ($Q^2 = 6.0$ GeV$^2$).}
\label{fig0}
\end{figure}

In the nucleus rest frame we can consider the interaction between a
virtual colorless hard probe and the nucleus via the gluon pair
($gg$) component of the virtual probe. The interaction of the
dipole with the color field of the nucleus will clearly depend on
the its size. If the separation of the $gg$ pair is very
small (smaller than the mean separation of the partons), the color
field of the dipole will be effectively screened and the nucleus
will be essentially transparent to the dipole. At large dipole
sizes, the color field of the dipole is large and it interacts
strongly with the target  and is sensitive both to its structure
and size. More generally, when the parton density is such that the
nucleus becomes black and the interaction probability is unity,
the dipole cross section saturates and  the gluon distribution
becomes proportional to the virtuality of the probe $Q^2$. In the
infinite momentum frame, this picture is equivalent to a situation
in which the individual partons become so close that they have a
significant probability of interacting with each other before
interaction with the probe. Such interactions lead, for instance,
to two $\rightarrow$ one branchings and hence a reduction in the
gluon distribution. These properties were considered in Ref.
\cite{ayala1}, where the rescatterings of the gluon pair inside
the nucleus were estimated using the Glauber-Mueller approach,
resulting  that the nuclear gluon distribution is given by
\begin{eqnarray}
xG_A(x,Q^2) = \frac{2R_A^2}{\pi^2}\int_x^1
\frac{dx^{\prime}}{x^{\prime}}
\int_{\frac{1}{Q^2}}^{\frac{1}{Q_0^2}} \frac{d^2r_t}{\pi r_t^4}
\{C + ln(\kappa_G(x^{\prime}, r_t^2)) + E_1(\kappa_G(x^{\prime},
r_t^2))\} \label{master}
\end{eqnarray}
where $C$ is the Euler constant,  $E_1$ is the exponential
function, the function  $\kappa_G(x, r_t^2) = (3
\alpha_s\,A/2R_A^2)\,\pi\,r_t^2\,
 xG_N(x,\frac{1}{r_t^2})$,
 $A$ is the number of nucleons in a nucleus and $R_A^2$ is the mean nuclear radius.
The limit of low densities is characterized by $\kappa_G \ll 1$, while for high parton  
densities $\kappa_G \gg 1$. The transition line between these two regimes can be obtained 
assuming $\kappa_G = 1$ \cite{plb96}, which allow us to estimate the saturation momentum 
scale $Q_s^2$. One of the shortcomings of this approach is that the Glauber-Mueller 
formula does not contemplate any nuclear effect in the nonperturbative initial condition 
for the gluon distribution. The antishadowing and EMC effects present at larger values  
of $x$ are also disregarded in this approach. In order to improve this approach, it was 
proposed in Ref.\cite{ayavic} a modification in the expression (\ref{master}) which 
includes the full DGLAP kernel in parton evolution. Basically,  to calculate the nuclear
gluon distribution  the following procedure was proposed:
\begin{eqnarray}
xG_A(x,Q^2) & = & (1/A)xG_A(x,Q^2)[GM] - (1/A)xG_A(x,Q^2)[DLA]
+ \, xG_A(x,Q^2)[EKS] \,\,\,, \label{proc}
\end{eqnarray}
where $xG_A(x,Q^2)[GM]$ represents the Glauber-Mueller nuclear gluon distribution given by 
the expression (\ref{master}) and $xG_A(x,Q^2)[DLA]$ is the DGLAP (DLA) prediction for the 
nuclear gluon distribution, which corresponds to the first term of expression (\ref{master}) 
when expanded in powers of $\kappa_G$. The last term in expression (\ref{proc}), $xG_A(x,Q^2)[EKS]$, 
is the gluon distribution solution of the DGLAP equation as proposed in Ref. \cite{eks}, 
where the initial conditions of the parton evolution were chosen in such a way to
describe the nuclear effects in DIS and Drell-Yan fixed nuclear target data. 
As it was discussed in detail in Ref.\cite{ayavic}, the parameterization of the nuclear effects 
in the gluon distribution, given by the ratio $R_g (x,Q^2) = xG_A(x,Q^2)/xG_N(x, Q^2)$ in 
the EKS procedure, does not include the perturbative high density
effects  at small values of $x$ associated to the QCD dynamics in this
regime.
Thus, the procedure  presented in equation (\ref{proc}) includes the full DGLAP evolution equation 
in all kinematic region, taking into account the nuclear effects in the present fixed target data
and   high density effects in the parton evolution at small $x$ in the
perturbative regime.
A similar procedure can be implemented to obtain the nuclear quark distribution \cite{marcos}.
In Fig. \ref{fig0} we present the predictions
for the $x$-dependence of nuclear ratios $R_g$ and $R_q = xq_A/xq_N$
considering the high density effects. The predictions of the EKS
parameterization are also shown for comparison. Two points must be
emphasized: (a) at  small values of $x$ the difference between the AG
and EKS predictions are very large and  merits our consideration;
(b) the effects for  gluons  are larger than for quarks. This last result has
strong implications in the calculations of the ratio between the $J/\Psi$
and Drell-Yan cross sections, as will be demonstrated in that follows.
A comment is in order here. Currently, there is a large uncertainty related to
the behavior of the nuclear parton distributions at small values of $x$. As
discussed in Ref. \cite{armestocapella}, the distinct models present in the
literature agree within 15 $\%$ for $x \approx 0.01$, where experimental data
exist, while they differ up to $60 \%$ in its predictions for $R_{F_2} = F_2^A/F_2^N$
at $x = 10^{-5}$. Our  prediction for this ratio using the AG parameterization is identical to
the EKS results for large $x$ and very similar to the eikonal one obtained
in Ref. \cite{armestocapella}. Therefore, we believe that the
approach used here can be considered as a suitable  phenomenological method 
for the inclusion of the high density effects in the nuclear cross sections 
at RHIC and LHC energies. Clearly, for higher values of parton density, a more 
general framework should be considered for the calculations of the cross 
sections (See e.g. Ref. \cite{gelis}).

% An alternative formulation is provided by the light-cone dipole
%pproach \cite{kopeliovich}. In this approach, the $c\bar{c}$ is
%produced over large longitudinal distances, which can exceeds the
%nuclear radius, in high energies (or low $x$) processes. This large
%enght scale leads to pronounced nuclear effects and it affects the
%QCD fatorization, due to coherence phenomena. Besides, in this approach
%the gluon shadowing is calculated as much stronger in charmonium production
%than in open charm production or DIS. Recently, a comparison between
%the predictions of the parton model and the dipole approach for heavy
%quark hadro-production was presented \cite{raufpeng}. In that work,
%one shows the equivalence between the two approaches for $pp$
%processes. In our present work, we choose the more conservative
%approach for treatment of nuclear and saturation effects.

\section{Energy Dependence for $J/\Psi$ Production}
\label{firstresult}

\begin{figure}[t]
%\begin{tabular}{c c}
\psfig{file=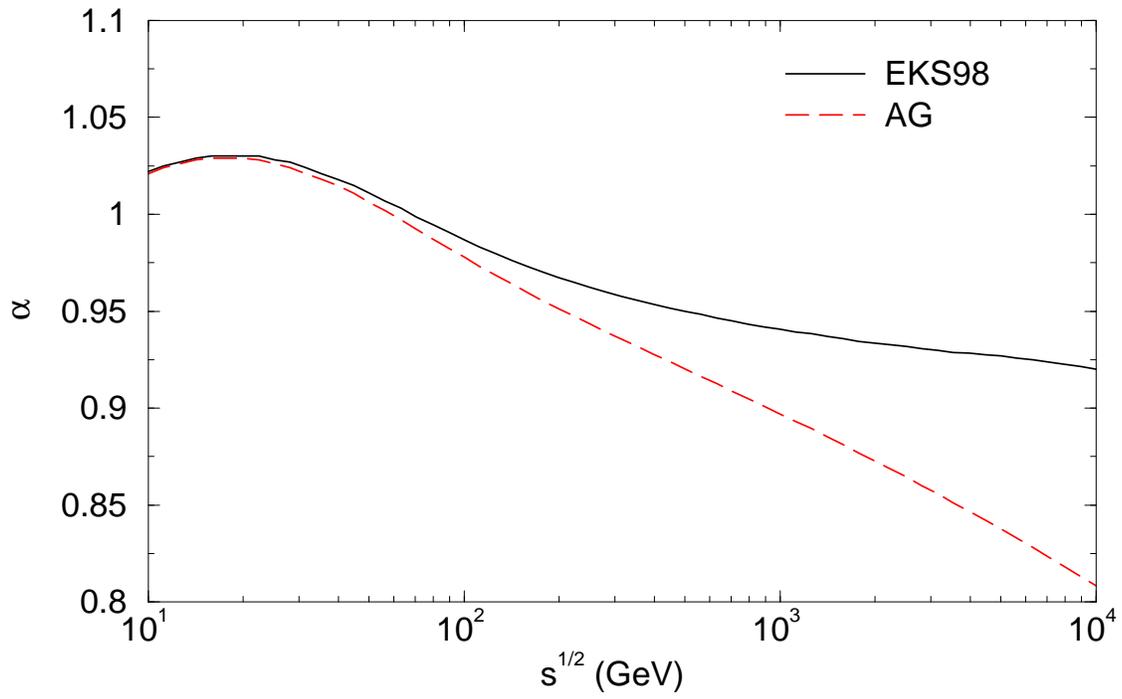,width=150mm}
%\psfig{file=bethaexpXenergy.eps,width=100mm}
%\end{tabular}
\caption{ Energy dependence of the effective
exponent $\alpha$ for $J/\Psi$ production in $pA$  processes. }
\label{fig1}
\end{figure}

In order to investigate the medium dependence of the $J/\Psi$ production 
cross section due to high density effects, we will follow the usual procedure to
describe the experimental data on nuclear effects in the hadronic
quarkonium production \cite{alfass}, where the atomic mass number
$A$ dependence is parameterized by $\sigma_{pA} = \sigma_{pN}
\times A^{\alpha}$. Here, $\sigma_{pA}$ and $\sigma_{pN}$ are the
particle production cross sections in proton-nucleus and
proton-nucleon interactions, respectively. If the particle
production is not modified by the presence of nuclear matter, then
$\alpha = 1$. In Fig. \ref{fig1} we present the effective exponent $\alpha$ as a
function of the c.m. energy $s^{1/2}$, where we have calculated the $J/\Psi$
cross section using the EKS and AG parameterizations as input in our calculations.
In both cases, we assume that the nucleon parton distributions are given by the
GRV94(LO) set \cite{grv94}.
 The general behavior of the exponent can be understood
as follows: when the integrations in Eq. (\ref{cross}) are taken,
the nuclear gluon distribution are evaluated in the $x_2$ interval
given by $\tau \, < \, x_2 \, < \, \sqrt{\tau} $. Thus, when the
energy grows, the $x_2$ interval goes to the small $x$ region. For
$J/\Psi$  production, for example, the antishadowing region
dominates the integration for energies smaller than 80
GeV. For larger  values of energy,
the suppression is sizeable and the effective exponent is smaller
than 1. In general grounds, we have that the exponent observed in 
the EKS prediction almost saturate and it is associated to the
behavior of the ratio $R_g$ at small $x$, while the presence of
the high density effects (nonsaturation of $R_g$) implies a large
reduction of the $pA$ cross section when compared with DGLAP-EKS
description of nuclear effects. Therefore, we believe that the
analysis of the effective exponent in $pA$ processes can be useful
to make clear the high density effects.

\begin{figure}[t]
%\begin{tabular}{c c}
\psfig{file=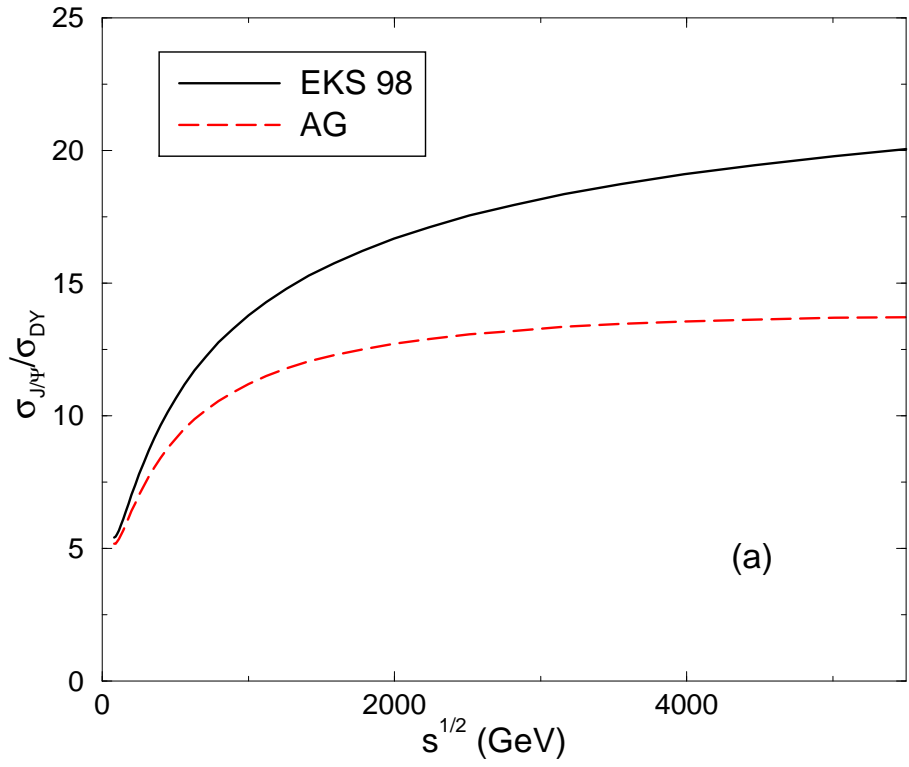,width=120mm}
\psfig{file=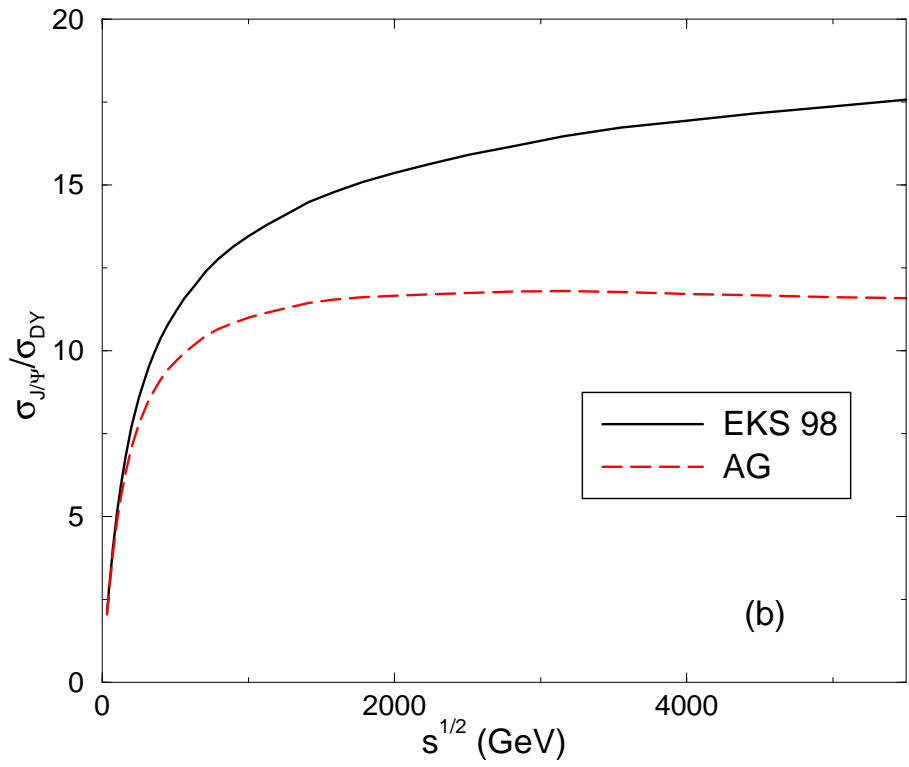,width=120mm}
%\end{tabular}
\caption{ Energy dependence of the ratio between the
$J/\Psi$ and DY cross sections for (a) $pA$ and (b) $AA$ collisions.
}
\label{fig2}
\end{figure}

In general, the experimental analysis of the  $J/\Psi$ suppression
\cite{abreu} consider as baseline the study of the ratio between
$J/\Psi$ and Drell-Yan (DY) cross sections. However, to make the
suppression of this ratio an unambiguous test, we must consider
all possible effects present in the production cross sections. In
particular, we cannot disregard  the high density effects in the
nuclear quark distributions and the larger magnitude of these
effects in the nuclear gluon distribution. In Fig. \ref{fig2} we
present our results for the energy dependence of the  ratio
$\sigma_{J/\Psi}/\sigma_{DY}$ for $pA$ (top panel) and $AA$
processes (bottom panel). For the calculation of the Drell-Yan
cross section we include the high density effects in the nuclear
quark distributions \cite{marcos}. As these effects are larger in the gluon case,
the ratio will be strongly modified when compared to the
calculations without the inclusion of the high density effects. We
verify that this modification is already present in $pA$ process
at RHIC energy, and is almost $40 \%$ for LHC energies, which
implies that this process could be used to identify the presence
and estimate the magnitude of the high density effects. Moreover,
we predict a  large suppression of the quarkonium production  in
$AA$ processes, associated  with the presence of these effects,
independently of the production  of a quark-gluon plasma. As the
additional suppression of the charmonium production rate predicted
here is associated to the distinct dependence of the $J/\Psi$ and
Drell-Yan cross sections in  the nuclear parton distributions, one
alternative analysis is to use as baseline the ratio $\sigma_{J
/\Psi}/\sigma_{Q\bar{Q}}$ between the bound and open states cross
section \cite{satzsridar}. Since both processes similarly depend
on the  nuclear gluon distribution this ratio is weakly dependent
on the inclusion of the high density effects \cite{disluiz}.

It is important to emphasize that the description of the
$J/\Psi$ production using collinear factorization at RHIC and LHC
is still an open question. At large energies the replacement of
the collinear factorization by the $k_{\perp}$-factorization
formalism is expected  (For a recent review see, e.g., Ref.
\cite{ktfatorization}). Moreover, coherence effects may lead
to breakdown the collinear factorization \cite{kopeliovich},
implying that dynamical effects, such as shadowing, may become
process dependent. In this work, we choose a more conservative
approach for treatment of the $J/\Psi$ production. We assume as
valid the collinear factorization and, in particular, the color
evaporation model, and consider that the high density effects only
modify the nuclear parton distributions. Moreover, we assume that
initial and final state effects can be separated and consider
different approaches for these distinct phases. The current
experimental data for particle production at large transverse
momentum in $\sqrt{s} = 200$ GeV  has demonstrated that this
approach is valid at least for RHIC energies.  We believe that our
analysis can be considered as a lower bound for the magnitude of
the high density effects in $J/\Psi$ production, since the
Glauber-Mueller approach considered here is one of the limits of
the color glass condensate for intermediate values of the parton
density. Clearly, for higher values of energy, a more general
framework should be considered in the calculations of the
quarkonium cross section.

\section{Final State Interactions}
\label{FSeffect}

In the later section we have demonstrated that the inclusion of
the high density effects implies an additional suppression of the
$J/\Psi$ production rate. However, a complete description of
nuclear $J/\Psi$ production requires a proper treatment of both
the initial state modifications of the parton distributions and
the final state interactions of the produced $c\overline{c}$
pairs. Therefore, in order to obtain a more realistic prediction
for the $J/\Psi$ suppression in heavy ion collisions, we must
include the final state interactions. Here, we will consider the
model recently proposed in Ref. \cite{qvz} (See also Ref.
\cite{chaud}), where the CEM is generalized to include the
multiple scatterings between the $c\overline{c}$ and soft gluons
as the pair cross  the nuclear medium.  In this model, the
$c\bar{c}$ is produced with a separation smaller than the $J/\Psi$
radius. Besides, the energy exchanges in the collisions are
greater than the nonperturbative momentum scales in the $J/\Psi$
wave function. Following this, the meson is unlikely to be formed
at the collision point and the transformation from a pre-resonant
$c\bar{c}$ into a physical $J/\Psi$ may occur over several Fermi
\cite{brodskymueller}. During this transformation, the produced
$c\bar{c}$ pair interacts via multiple scatterings with the
spectators of the collision \cite{qvz_npa698}. As a consequence of
the multiple scattering, the square of the relative momentum $q^2$ between the
$c\bar{c}$ pair is increased, and some of the $c\bar{c}$ pairs could
gain enough $q^2$ to reach the threshold for open charm production. As
a result, the cross sections for $J/\Psi$ production is reduced in
comparison with nucleon-nucleon collisions.

In the QVZ model \cite{qvz}, the transition probability $F_{c\bar{c} \rightarrow J/\Psi}(q^2)$,
present in Eq. (\ref{cross}), is parameterized by
\begin{eqnarray}
F_{c\bar c\to {\rm J}/\psi}(q^2)
&=&N_{{\rm J}/\psi}\theta(q^2)\theta(4{m_D}^2-4m_c^2-q^2)
\left (1-\frac{q^2}{4{m_D}^2-4m_c^2}
\right )^{\alpha_F} \, ,
\label{fgene}
\end{eqnarray}
where  the gluon radiation effect is simulated by the parameter
$\alpha_F>0$, which  put  larger weight to the smaller $q^2$.
Furthermore, the QVZ model additionally assumes that the multiple
scattering of the pair in the nuclear medium would increase the
relative square momentum $q^2$ of the pair.
The effect of coherent multiple scattering
in the perturbative QCD calculation
may be represented by shifting of the relative momentum
in the transition probability as \cite{qvz}
\begin{equation}
F_{c\bar c\to {\rm J}/\psi}(\bar q^2)
=F_{c\bar c\to {\rm J}/\psi}(q^2+\varepsilon^2 L),
\label{shift}
\end{equation}
where $L$ is the effective length of the nuclear medium in the $AB$ 
collisions and $\epsilon^2$ is the squared relative momentum gained 
by the produced $c\overline{c}$ pair per unit length. The above behavior
implies that  for a large enough $L$, such that for $\bar
q^2>4{m_D}^2-4m_c^2$, the transition probability essentially
vanishes due to the existence of the open charm threshold. This
gives rise to a much stronger suppression than the exponential one
following from the Glauber model \cite{fujii}. In principle, the
uniform shift of the relative momentum may be obtained in a
perturbative QCD calculation by summing up all twist contributions
at the leading order in the strong coupling constant $\alpha_s$
and the nuclear size \cite{fries}. As shown in ref. \cite{qvz},
the model describes all observed $J/\Psi$ suppression data in
hadron-nucleus and nucleus-nucleus collisions if the parameters
$\alpha_F$ and $\epsilon$ are fitted. Moreover, the model is
weakly dependent of the value of $\alpha_F$, which is not true for
$\epsilon^2$. It is important to emphasize that the authors have
disregarded the effects in the nuclear parton distributions in the
calculations. Below, we demonstrate that the inclusion of these
effects is  important for the SPS energies and should be included
in the future predictions for RHIC and LHC energies.

\section{Results and Conclusions}
\label{comments}

In the later sections we have presented the model to treat the medium
effects in nuclear collisions. In Section \ref{ISeffect}, we have discussed the initial
state effects, and in the Section \ref{FSeffect}, we have presented a
model to treat the final state effects, and its implications for the
$J/\Psi$ suppression. In this section, we present our results for $J/\Psi$ production, when
these effects are considered. Since we are interested in the magnitude of the suppression due to
high density effects, we present the results for total cross
sections for charmonium productions, thought  only the small
$x_{F}$ region will be accessible at RHIC and LHC.

\begin{figure}[t]
\centerline{\psfig{file=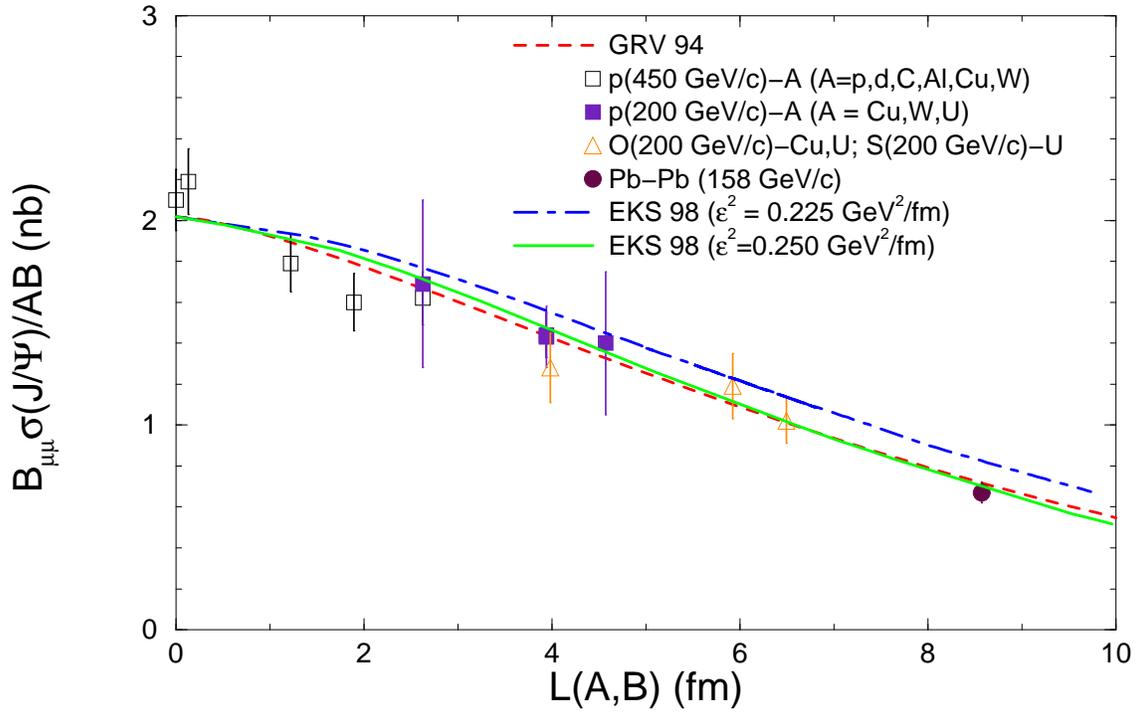,width=150mm}} \caption{Total $J/\Psi$ cross sections with the
branching ratios to $\mu^{+} \mu^{-}$ in $pA$ and $AA$ collisions, as
a function of the effective nuclear lenght $L(A,B)$.}
\label{fig3}
\end{figure}

In Fig. \ref{fig3} we show our results for the medium length dependence
of the $J/\Psi$ cross section. The effective length $L(A,B)$, as well
as the experimental data, are taken from the Ref. \cite{abreu}, with
all data rescaled to $P_{\mbox{beam}} = 200$ GeV. Following
Ref. \cite{qvz}, we assume $K\,N_{J/\Psi} = 0.458$. The dashed line is
our prediction obtained using the GRV94 parameterization and, as made in Refs.
\cite{qvz,chaud},  we assume $\epsilon^2 = 0.225$ GeV$^2$/fm. In this 
case we are disregarding any nuclear effects present in the parton 
distributions. If  these nuclear effects are considered, we obtain the 
dash-dotted curve, where we kept $\epsilon^2 = 0.225$ GeV$^2$/fm and 
used the EKS parameterization. It can be seen that if these effects are 
included the model fails to describe the $Pb-Pb$ data. This behavior is 
associated to the antishadowing effect present in the EKS nuclear parton 
distribution, which implies an enhancement of the cross section and
more scatterings between the $c\bar{c}$ pair and the medium are necessary 
to describe the data. This situation can be improved using the fact that 
$\epsilon^2$, the gained momentum per unit length, is a free parameter 
of the model. In the solid line we show our prediction using 
$\epsilon^2 = 0.250$ GeV$^2$/fm, which improves the description. This 
modification in the parameter demonstrates that the inclusion of the 
medium effects are important for SPS energies. It is important to emphasize 
that our conclusions are unchanged if we use for comparison the latest 
NA50 data on $J/\Psi$ production in $pA$ and $AA$ collisions \cite{NA50last}. 
As in the kinematical region of the experimental data the EKS and AG 
predictions for the nuclear parton distributions are identical, our 
predictions including the high density effects are not shown in the figure.

\begin{figure}[t]
%\begin{tabular}{c c}
\psfig{file=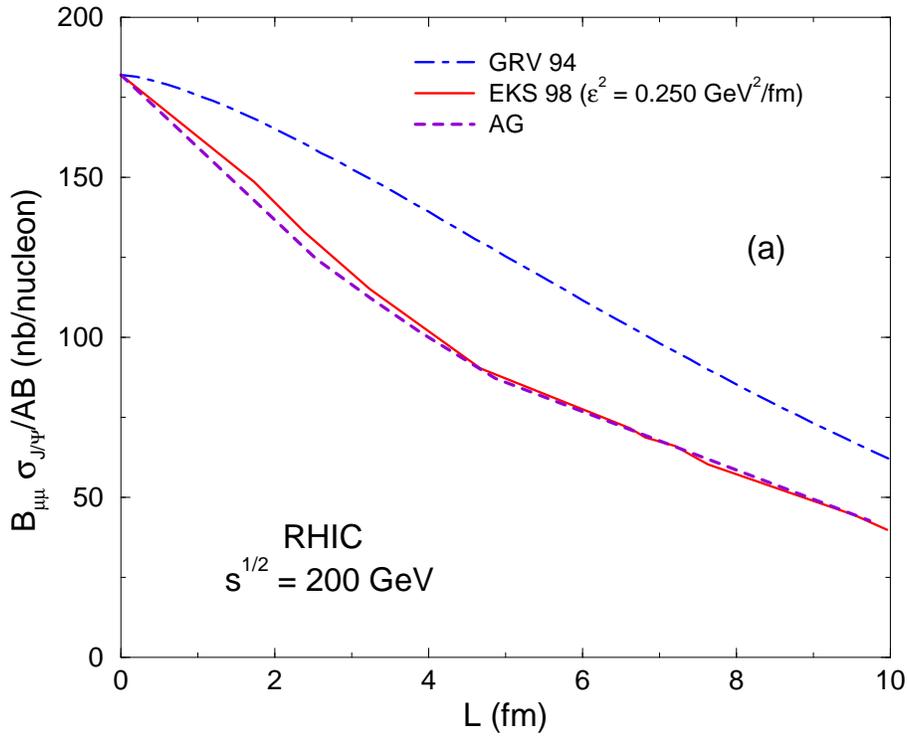,width=120mm}
\psfig{file=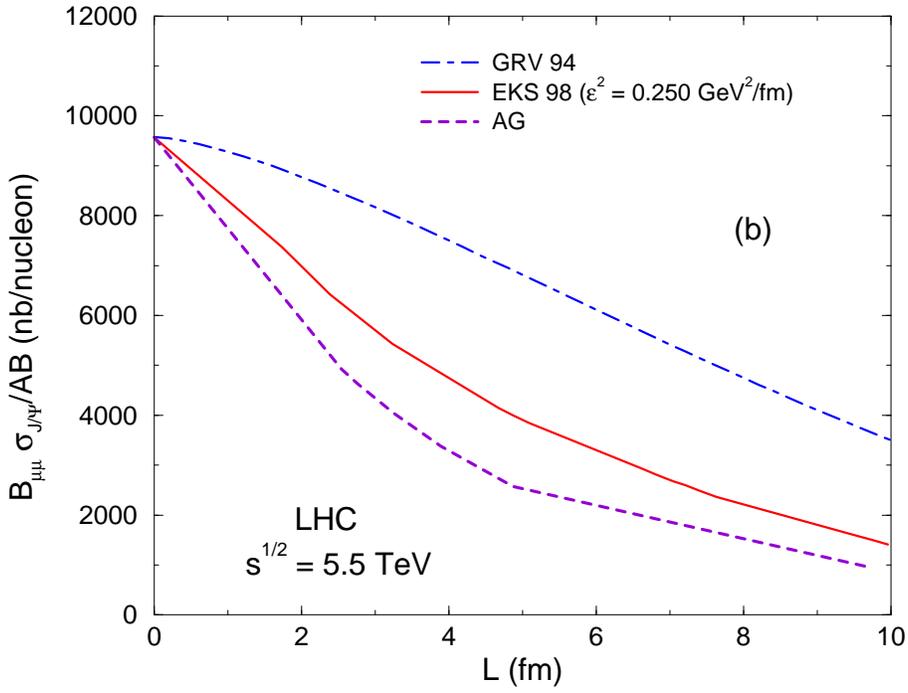,width=120mm}
%\end{tabular}
\caption{The same as Fig. \ref{fig3}, for (a)
RHIC and (b) LHC energies. }
\label{fig4}
\end{figure}

As we have determined a new value for the parameter $\epsilon^2$
when the initial state medium effects are considered, we can use it to predict the
nuclear $J/\Psi$ production at higher energies. We assume that a QGP
is not formed in these collisions. Consequently, 
our predictions are a
lower bound for the  $J/\Psi$ suppression, since if this new state of
matter is formed, another mechanism of suppression (deconfinement) not
considered here is also present.
 In Fig. \ref{fig4} we plot the 
our predictions for total cross sections for RHIC and LHC energies, 
using the GRV, EKS and AG parameterizations as input in our calculations. We
can see by the comparison between the GRV and EKS curve that for the RHIC
and LHC energies the medium effects present in the initial state cannot
be disregarded, with the inclusion of these effects implying an additional
$J/\Psi$ suppression. Furthermore, for the kinematical region of RHIC
the inclusion of the high density effects does not significantly modify the
behavior of this observable. However, for LHC energies, we predict that
these effects strongly reduces the $J/\Psi$ production rate. This effect
is associated to the dominance of the small-$x$ behavior of the nuclear
parton distributions at high energies.

A complete description of nuclear $J/\Psi$ production requires a
proper treatment of both the initial state modifications of the
parton distributions and the final state interactions of the
produced $c\overline{c}$ pairs. In this paper we have considered
the inclusion of the high density effects in the parton
distributions and estimated the $J/\Psi$ cross section in $pA$ and
$AA$ processes assuming that the collinear  factorization is still
valid. Our results have demonstrated that these effects cannot be
disregarded in the kinematical regions of the future colliders. In
particular, we have shown that an additional $J/\Psi$ suppression
is expected if these effects are included in the calculations. As
in  $AA$ collisions this suppression can also be associated to
other sources, as for instance,  comover interactions, percolation
deconfinement or quark-gluon plasma formation, an experimental
analysis of the $J/\Psi$ production in $pA$ processes at high
energies  is fundamental to constraint the hadronic effects. In
particular, our results  demonstrate that the high density effects
will significantly contribute in the $J/\Psi$ production in these
processes, which implies that a careful disentangling of initial
and final state effects is required before the clear
identification of a new state of matter.

\section*{Acknowledgments}
 V.P.G. and  L.F.M. are very grateful to  A. Ayala Filho,
M. A. Betemps and M. V. T. Machado for illuminating discussions on the subject.
This work was partially financed by CNPq  and FAPERGS, BRAZIL.

\newpage

\newpage

\newpage

\newpage

\newpage

\end{document}